\documentstyle[11pt,aaspp4]{article}

\newcommand{\Teff}{$T\sb{\rm eff}$}
\newcommand{\logg}{$\rm log~{\it g}$}

\newcommand{\Lalpha}{\hbox{L$\alpha$}}
\newcommand{\Lbeta}{\hbox{L$\beta$}}
\newcommand{\Hbeta}{\hbox{H$\beta$}}
\newcommand{\Lgamma}{\hbox{L$\gamma$}}

\lefthead{Koester et al.}
\righthead{Quasi-molecular satellites in Lyman Beta}

\begin{document}

\title{QUASI-MOLECULAR SATELLITES OF LYMAN BETA IN THE SPECTRUM OF THE
DA WHITE DWARF WOLF~1346\footnotemark[1]}

\footnotetext[1]{Based on observations obtained with
    the Hopkins Ultraviolet Telescope and optical observations obtained at
    the DSZA Calar Alto}

\author{D. Koester}
\affil{
   Institut f\"ur Astronomie und
   Astrophysik, Universit\"at Kiel, D--24098 Kiel, Germany.
   E-mail: koester@astrophysik.uni-kiel.d400.de}

\author{D. S. Finley\altaffilmark{2}}
\affil{
   Eureka Scientific, Inc.
   2452 Delmer St. Suite 100,
   Oakland, CA  94602}
\altaffiltext{2}{
   Center for EUV Astrophysics, University of California, Berkeley, CA  94720.
   E-mail: david@cea.berkeley.edu}

\author{N. F. Allard\altaffilmark{3}}
\affil{
   Observatoire de Paris-Meudon, F--92195 Meudon, France}
\altaffiltext{3}{
   CNRS Institut d'Astrophysique, F-75014 Paris, France.
   E-mail: allard@iap.fr}

\author{J. W. Kruk}
\affil{
   Johns Hopkins University, Center for
   Astrophysical Sci., Baltimore, MD 21218.
   E-mail: kruk@hut4.pha.jhu.edu}

   \and

\author{R. A. Kimble}
\affil{
   NASA/GSFC, Greenbelt, MD 20771.
   E-mail: kimble@stars.gsfc.nasa.gov}

\begin{abstract}

We present new FUV/UV observations of the DA white dwarf Wolf~1346
obtained with the Hopkins Ultraviolet Telescope. The atmospheric
parameters of this object are estimated from a fit of model
atmospheres to several optical spectra to be \Teff~ = 20000~K, \logg~
= 7.90. From the optical spectrum this star is a normal DA without any
indications for chemical elements other than hydrogen. The hydrogen
line \Lbeta, however, shows a very unusual shape, with a steep red
wing and two absorption features on this wing. The shape is
reminiscent of the effects of quasi-molecular line broadening, as
observed in \Lalpha\ in cooler DA white dwarfs. We show that this is
indeed the correct explanation, by identifying 4 quasi-molecular
satellites caused through perturbations by the H$^+$ ion (H$_2^+$
quasi-molecule). The steep red wing is caused by the exponential
decline of the line profile beyond the satellite most distant from the
line center at 1078~\AA.
\end{abstract}


\section{Introduction}

Throughout the recent history of astronomy it has happened repeatedly
that the opening of a new observing window into space has lead to
discoveries that were not expected.  Just in the field of white dwarfs
the systematic exploration of the UV window has brought unexpected
surprises. We mention just a few out of a large number: the discovery
of metal features in hot DA, very strong carbon lines in DC white
dwarfs, which show no features at all in the optical, and
quasi-molecular satellites in cool DA, which were expected to show
just a simple Stark broadened \Lalpha\ line.

One of the least explored spectral windows remains the FUV region,
roughly defined as 912 - 1215 \AA, the region from the Lyman edge to
\Lalpha. While the interstellar matter is quite transparent at these
wavelengths, as opposed to the region below the Lyman edge, progress
has been slow due to technological difficulties in the production of
mirrors and gratings. This situation is improving recently with
projects like ORFEUS (Orbiting Retrievable Far and Extreme Ultraviolet
Spectrograph; Grewing et al. 1991; Hurwitz \& Bowyer 1991) and HUT
(Hopkins Ultraviolet Telescope, Davidsen et al. 1992).  In this paper
we report on an observation of the bright DA white dwarf Wolf~1346
(WD2032+248), obtained with the HUT instrument on a flight in
1995. The observation was part of an Astro-2 Guest Investigator
program (Finley, Kimble, and Koester) aimed at studying the Stark
broadening of the higher Lyman lines in DA. While several hotter DA
show the whole Lyman spectrum compatible with symmetrical Stark
broadened profiles without any unexplained features, Wolf~1346 at
20000~K has a \Lbeta\ line with a strong asymmetry, a very steep red
wing, and absorption features in the wing near 1060 and 1078~\AA. We
demonstrate below that all these features are due to quasi-molecular
H$_2^+$ absorption (or broadening of \Lbeta\ by protons as
perturbers), very similar in nature to the 1400~\AA\ feature observed
at lower temperatures in the red wing of \Lalpha.

\section{UV/FUV observations with the Hopkins Ultraviolet Telescope}

Our FUV spectra were obtained with the Hopkins Ultraviolet Telescope
(HUT), one of the instruments in the Astro-2 mission that was flown as
an attached space shuttle payload in March, 1995.  The initial
observation of Wolf~1346 was made on 5 March 1995.  The spectrum
contained wholly unexpected features in the Lyman lines that were so
unusual that concerns were raised that some anomaly may have occurred
during the observation.  Consequently, the real-time scheduling
capabilities of Astro-2 were utilized to reobserve Wolf~1346 on 14
March.  The two spectra were virtually identical, and no other white
dwarfs showed the unusual features, confirming that the features were
intrinsic to Wolf~1346.

The data reduction process included correcting for pulse pile-up;
subtracting dark counts, scattered light, and second order light;
flat-fielding; and correcting for time-dependent throughput changes
during the mission.  The pointing during the observations was such
that some light was lost due to the target being near the edge of the
aperture during much of the exposures.  Hence, an additional
normalization correction was made using the count rates obtained
during the times the target was fully within the aperture.  The
spectra were fluxed using the instrument sensitivity determined in
flight using observations of HZ~43 and a model computed with Koester's
atmosphere codes.  Finally, the background was subtracted using sky
exposures taken during target acquisition, and the two spectra were
coadded.  The data reduction process and in-flight calibration are
detailed in Kruk et al. (1995).
\begin{figure}[htbp]
\plotone{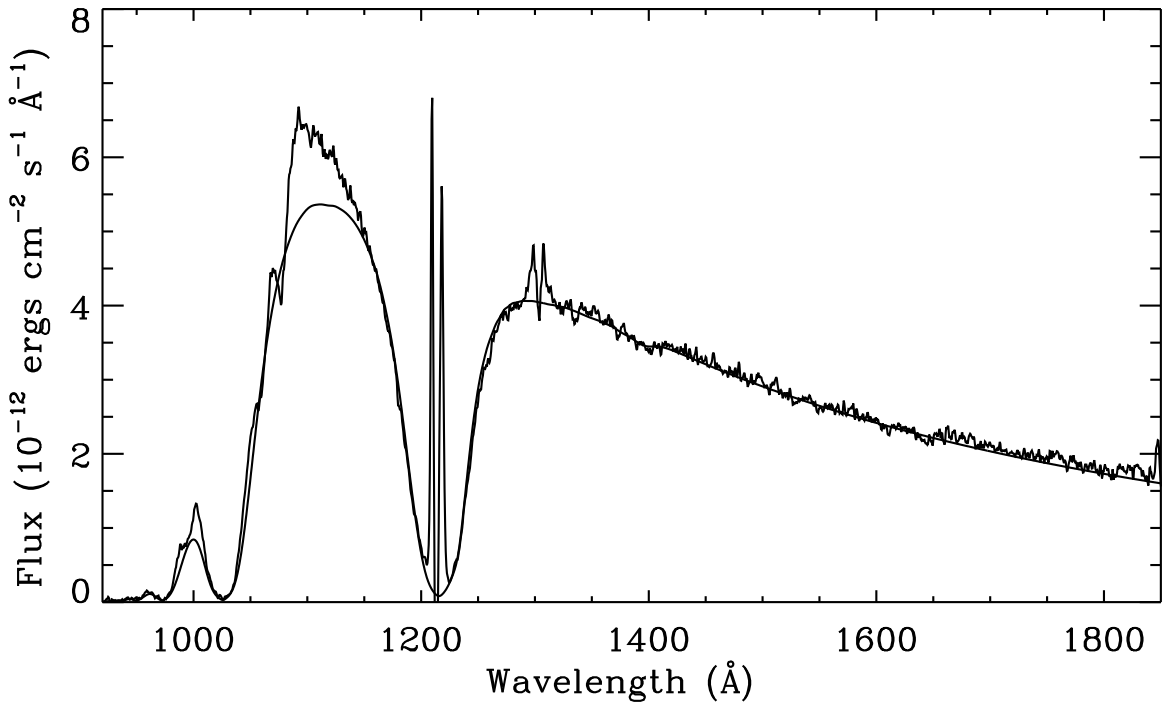}
\figcaption[]{Background-subtracted HUT spectrum of Wolf~1346, compared with a
theoretical spectrum using standard \Lbeta\ Stark broadening.
The \Lalpha\ and OI airglow lines at 1216 and 1304 \AA~ could not be
subtracted cleanly due to the decline in sensitivity in those lines
through the mission.
\label{hutspec}}
\end{figure}
The combined HUT spectrum is shown in Fig.~\ref{hutspec} and compared
with a model of our atmosphere grid, using standard Stark broadening.
The model was convolved with the instrument resolution (which varied
as a function of wavelength) for comparison with the observed
spectrum.  The spectrum was smoothed with a 2-pixel (1~\AA) FWHM
Gaussian, as was the model.  The model was scaled using the published
V magnitude of 11.53 (Cheselka et al. 1993) and the relation
$f_{\lambda} (5490 ~\rm{\AA}) ~=~ 3.61~10^{-9} / 10^{0.4 m_{\it{V}}}$
(Finley, Basri \& Bowyer 1990).  The model required scaling by an
additional factor of 1.03 to match the continuum flux between
1350~\AA~ and 1650~\AA.

The model does not reproduce the steep red wing of \Lbeta.  In
addition, there are two unexplained absorption features on the
observed wing at 1060 and 1078~\AA. On the red wing of \Lalpha\ a weak
indication of the 1400~\AA\ H$_2^+$ satellite is visible in both the
model and the observation.  The HUT effective area (Kruk et al. 1995)
is smooth in the 1060 -- 1080~\AA~ region.  Furthermore, the observed
spectra of the hotter DA that were observed agree quite well with the
models (Kruk et al. 1995), clearly demonstrating that the \Lbeta\
features are not calibration artifacts, but are instead intrinsic to
Wolf~1346.

\section{Optical observations and stellar parameters}

With a visual magnitude of 11.53, Wolf~1346 is a very bright white
dwarf and has therefore been included in a number of recent
analyses. Bergeron et al. (1992) give 19980/7.83 (\Teff/\logg), Finley
et al. (1996) 19960/7.83, and Kidder (1991) 20400/7.90. We have
obtained 3 spectra with high S/N at the 2.2m telescope of the DSAZ
observatory at Calar Alto in September 1995, and the analysis with a
set of model atmosphere grids gives 19500/7.96.
\begin{figure}[htbp]
\plotone{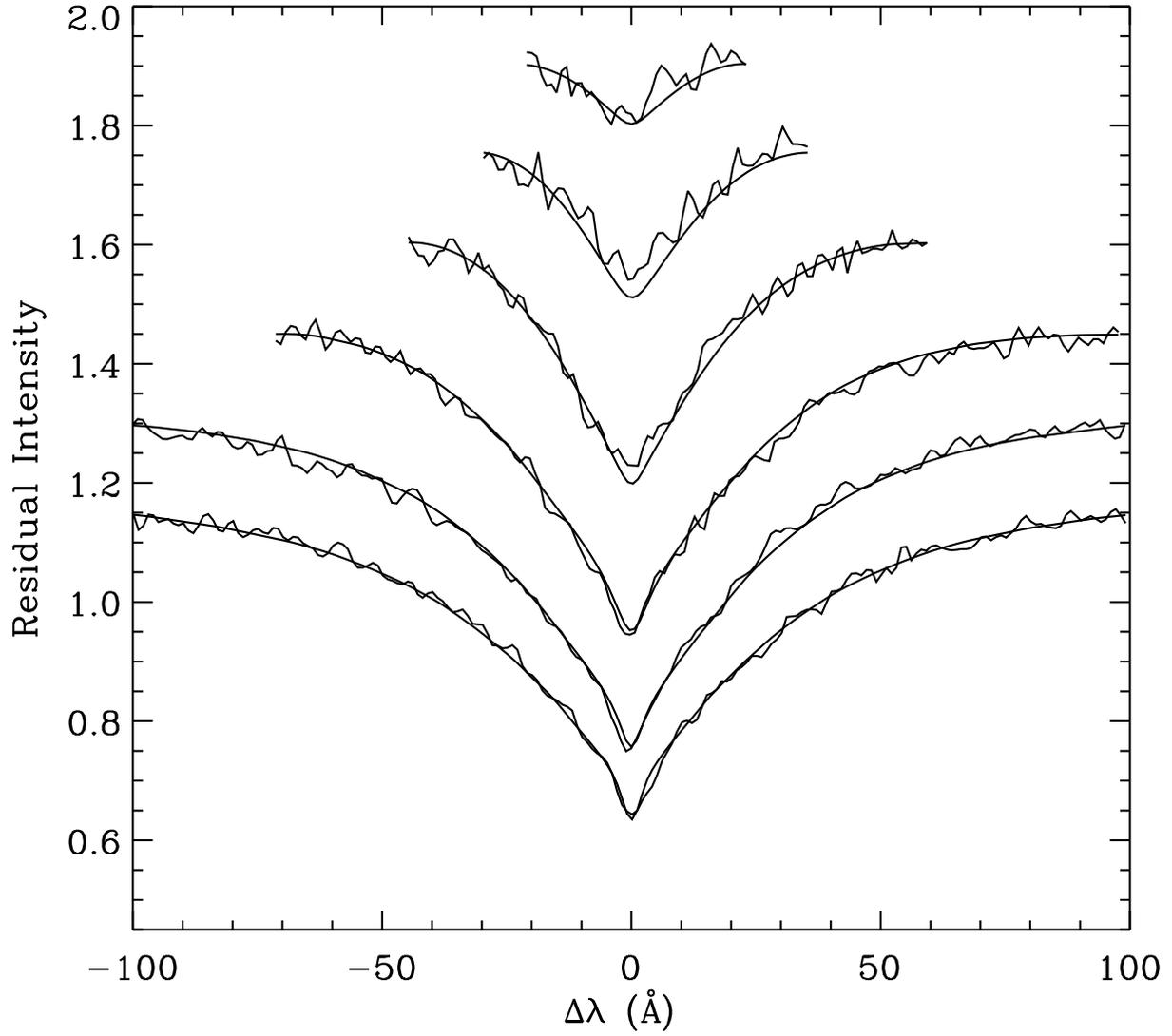}
\figcaption[]{Balmer line profiles of Wolf~1346 and the
best fitting model with \Teff~ = 19504~K, \logg~ = 7.96.
Lines shown are \Hbeta\ through H9.
\label{opt}}
\end{figure}
Fig.~\ref{opt} shows the Balmer line profiles from one of these
observations together with the best fit model.  The fitting procedure
included adjusting the models at continuum points to correct for the
non-perfect calibration of the observed spectrum, then simultaneously
fitting the \Hbeta\ -- H9 lines, including small sections of adjacent
continuum.  This means that the fit was determined essentially by the
Balmer line profiles only. The formal errors of our fits are very
small, comparable to those obtained by the other studies.  The slope
of the spectrum before adjustment of the continuum -- while not
totally reliable -- indicates that the temperature could perhaps be a
bit higher, as indicated by the other results. In any case, the
scatter of these determinations indicates that systematic errors are
probably higher than the formal errors determined by the least-squares
fits, and we adopt very conservatively the following values and ranges
for our further study: \Teff~ = 20000 $\pm$ 500~K, \logg~ = 7.90 $\pm$
0.10.

\section{Quasi-molecular line broadening of \Lbeta\ by protons}

In the model used for the description of this line broadening, the
interaction of the absorbing hydrogen atom in the ground state with
the perturbing proton is considered as the temporary formation of a
quasi-molecule, in this case the molecular ion H$_2^+$. The
perturbation of the energy levels is then given by the adiabatic
potential energy curves of this molecule.

The approach is based on the unified theory of Anderson \& Talman
(1956) and has in recent years been considerably improved through the
work of N. Allard and J. Kielkopf (Allard \& Kielkopf 1982, 1991;
Allard \& Koester 1992; Kielkopf \& Allard 1995). A comprehensive
recent review of these calculations is Allard et al. (1994).

In these papers the interest was concentrated on the line \Lalpha,
because previously unknown features in the IUE spectra of a number of
cool DA white dwarfs had been identified by Koester et al. (1985) and
Nelan \& Wegner (1985) as quasi-molecular satellites of \Lalpha\ due
to perturbations with protons and the neutral hydrogen atom.  The
application of the new profiles to IUE and HST/FOS spectra has
resulted in much improved fits for the UV spectra of cool DA white
dwarfs (e.g. Koester \& Allard 1993; Koester et al. 1994; Bergeron et
al.  1995) and to the identification of the famous 1600~\AA\ feature
in the $\lambda$ Bootis stars as due to the H$_2$ quasi-molecule
(Holweger et al. 1994).

It was therefore quite natural to suspect similar physical effects to
be responsible for the strange \Lbeta\ profile in Wolf~1346. In this
case, the situation is simplified by the fact that at the high
temperature of this object the only possible perturber is the
proton. We have therefore used essentially the same methods and
program codes as described in Allard et al. (1994) and applied them to
the transitions corresponding to \Lbeta, that is between those
molecular states that asymptotically (at large internuclear distance)
correspond to the first and third energy levels of the isolated
hydrogen atom.

The 2 states corresponding to n=1 and 12 states corresponding to n=3
are are listed by Ramaker \& Peek (1972, RP). From the transition
rules, and from the dipole moments calculated by RP, it is apparent
that there are 8 dipole-allowed transitions. We list these transitions
in Table~\ref{trans}, together with some relevant information that
will be discussed below.  The relative weights tabulated for the
individual transitions were determined from the Stark broadening
calculations of Underhill \& Waddell (1959).  The potential energy
curves for all these states are given by Madsen \& Peek (1971). 
\begin{table}
\caption[]{Allowed molecular transitions corresponding to the n=1 to
n=3 transitions in the isolated H atom. The first column is a running
number and the column labeled RP gives the numbers for the states as used
by RP.
The column labeled wt gives the relative weight.
The last column indicates the position of a satellite, if there
is one, in cm$^{-1}$ from the line center, and as absolute wavelength
position in \AA.}
\begin{flushleft}
\begin{tabular}{rrrrrrr}
\hline\noalign{\smallskip}
 No. & low & up & RP & wt & wing & $\Delta \omega $[cm$^{-1}] (\lambda$ [\AA])\\
\hline\noalign{\smallskip}
 1 & $1s\sigma_g$ & $4p\sigma_u$ & 1 - 5 & 1 & blue & --- \\
 2 & $2p\sigma_u$ & $3s\sigma_g$ & 1 - 5 & 1 & blue & --- \\
 3 & $1s\sigma_g$ & $3p\pi_u   $ & 1 - 8 & 2 & blue & --- \\
 4 & $2p\sigma_u$ & $4d\pi_g   $ & 1 - 8 & 2 & blue & --- \\
 5 & $1s\sigma_g$ & $6h\sigma_u$ & 1 - 7 & 1 & red  & -953 \quad (1036)\\
 6 & $2p\sigma_u$ & $5g\sigma_g$ & 1 - 7 & 1 & red  & -4696\quad (1078)\\
 7 & $1s\sigma_g$ & $4f\pi_u   $ & 1 - 9 & 2 & red  & -3124\quad (1060)\\
 8 & $2p\sigma_u$ & $5g\pi_g   $ & 1 - 9 & 2 & red  & -436 \quad (1030)\\
\hline\noalign{\smallskip}
\label{trans}
\end{tabular}
\end{flushleft}
\end{table}
\begin{figure}[htbp]
\plotone{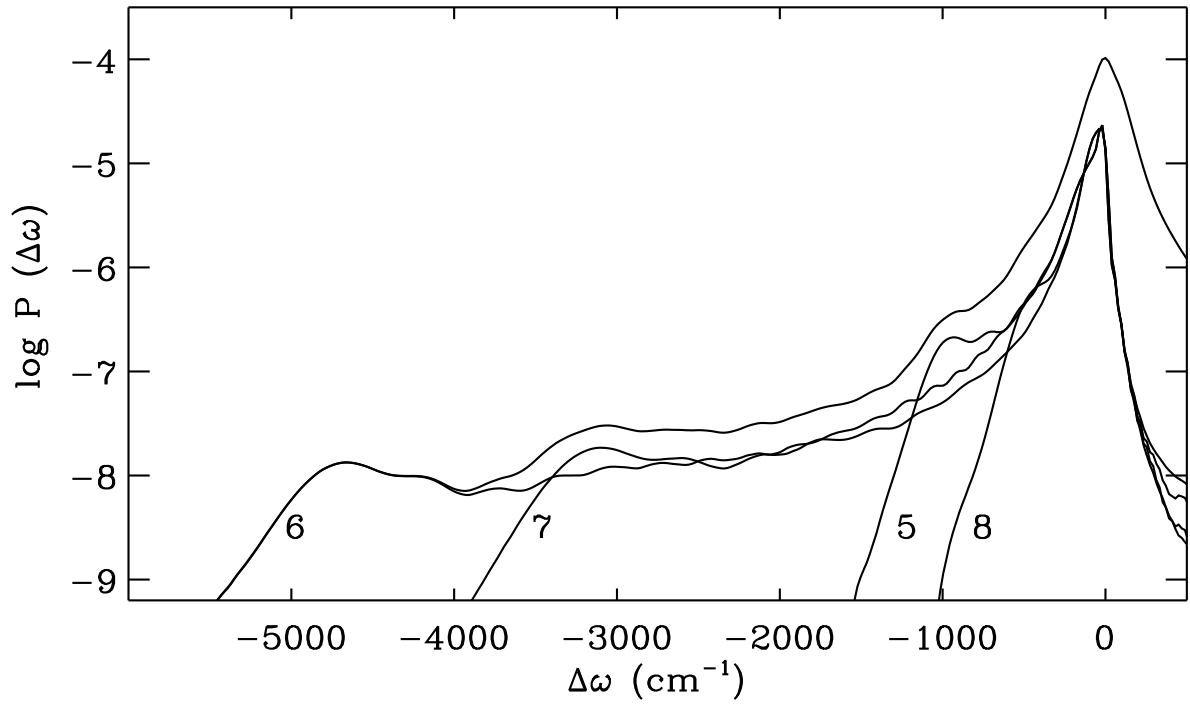}
\figcaption[]{Logarithmic line profile for \Lbeta\ as a function of
distance $\Delta\omega$ from the line center, showing the individual
contributions of the 4 transitions contributing to the red wing and the
total profile (upper line). The individual contributions are numbered;
the numbers correspond to the first column in Table 1. A temperature
of 20000~K and proton density of 10$^{16}$ cm$^{-3}$ was used for
this calculation. 
\label{redwing}}
\end{figure}
The numerical calculation of the correlation functions for all
transitions and the 8 individual contributions to the line profile
lead to the following results: each of the 4 transitions contributing
(at larger distance from the line center) to the red wing of \Lbeta\
shows a satellite feature; the positions are indicated in
Table~\ref{trans} as differences from the line center ($\Delta \omega$
in cm$^{-1}$) and also with their positions on the wavelength scale in
\AA. Even without the calculation of detailed synthetic spectra it is
clear that the 2 satellites farthest from the center correspond
exactly to the features seen in the HUT spectrum; the other two are
weaker and lost in the saturated line center. The blue wing of the
line shows no satellites, not even in the individual contributions of
the 4 transitions. We therefore show only the red wing, to be able to
show more of the interesting details, in Fig. \ref{redwing}.

Readers interested in obtaining line profile data should contact
N.~Allard or D.~Koester.

\section{Calculation of model atmospheres and synthetic spectra}

The \Lbeta\ line profiles have been incorporated in the LTE stellar
atmosphere codes for white dwarfs developed and improved by D.K. over
two decades. A rather dated description of these codes is in Koester
et al. (1979); a more up-to-date version will be given in Finley et
al. (1996). The profiles with the quasi-molecular satellites are used
consistently for the determination of the atmospheric structure
(``line blanketing'') as well as for the detailed spectrum synthesis.
The line blanketing is important for the \Lalpha\ satellites at lower
temperatures; the effect from \Lbeta\ is small.
The temperature dependence of the line profiles is small, and for the
\Lbeta\ line we have used only profiles calculated for a temperature
of 20000~K. The absorption coefficient per absorbing atom in the line
wing is roughly proportional to the perturber density, whereas the
shape of the satellite feature {\em relative} to the neighboring line
wing is independent of density. The satellite will therefore become
visible in a spectrum, whenever the line profile is visible out to the
position of the satellite. According to our preliminary model
calculations for \Lbeta\ this is the case in DA white dwarfs between
effective temperatures of about 16000 to 25000~K. This agrees with the
observational HUT result that all other DA observed are hotter than
30000~K.

\begin{figure}[htbp]
\plotone{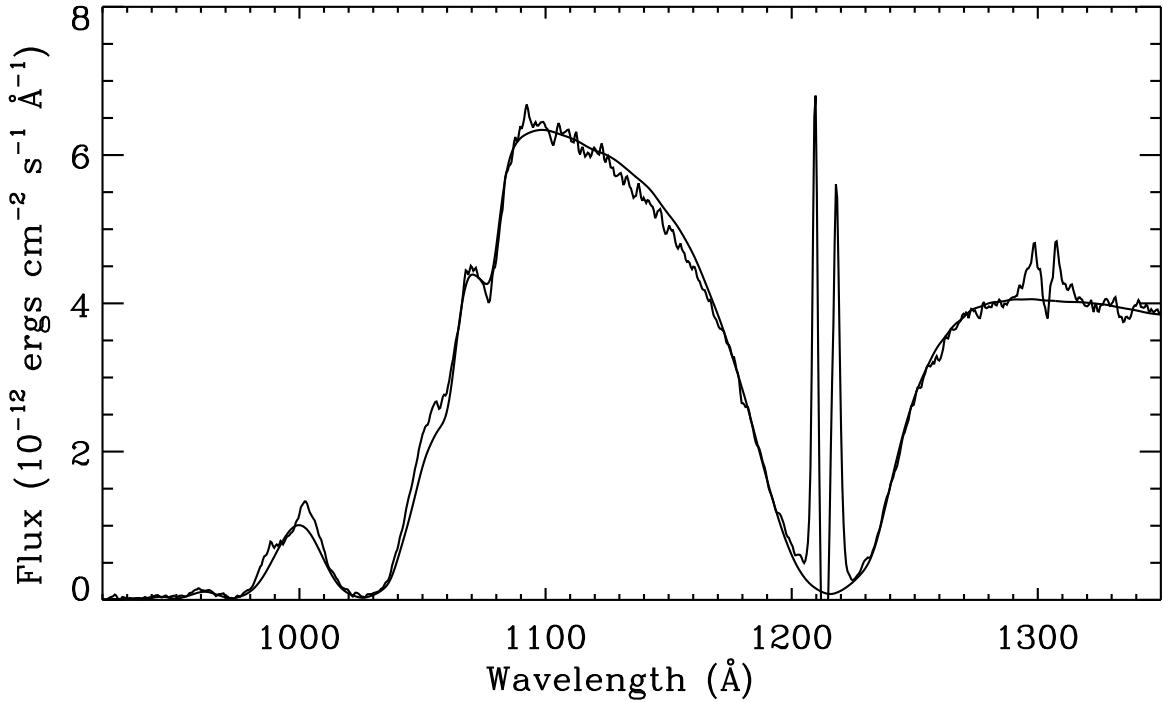}
\figcaption[]{\Lbeta\ and \Lalpha\ region of the background-subtracted
HUT spectrum compared with a theoretical model (smooth curve) with the
new \Lbeta\ broadening including the quasi-molecular satellites and
\Teff~ = 20000~K, \logg~ = 7.90.
\label{hutsat}}
\end{figure}
Fig.~\ref{hutsat} shows the result of this calculation compared to the
observation, using our adopted values for \Teff~ and \logg~
(20000/7.90).  The model was scaled using the V magnitude, with an
additional scale factor of 1.03 being applied to match the observed
continuum flux longward of \Lalpha.  A slightly hotter model (\Teff~ =
21000~K) did not give a good fit to the data.  The continuum flux for
the hotter model (scaled to V) was 10\% higher than observed and the
\Lalpha~ line was far too weak.

\section{Conclusions}

We have not made an effort to find the best fitting UV model, and the
fit is clearly not perfect, especially in the region between \Lalpha\
and \Lbeta. We have made some experiments and our conclusion is that
far wing absorption of \Lgamma\ and higher Lyman lines, which are
still calculated with standard Stark broadening, are part of the
problem. Fig.\ref{hutsat}, however, clearly proves, by the coincidence
of position and shape, that the two observed features near 1060 and
1078 \AA\ are indeed satellite features of \Lbeta, and that the steep
rise of the wing is caused by the exponential decline of the line
profile beyond the last satellite. Further detailed studies of the
line profiles of all Lyman lines will hopefully improve the
quantitative agreement in the future, whereas new observations of
hotter and cooler objects should establish the range where these
features are observable, and provide a challenge to experimental
physicists to measure these line profiles in laboratory plasmas.

\acknowledgments
N.F.A. thanks NATO for a Collaborative Research Grant (920167).
D.S.F. was supported by an Astro-2 Guest Investigator contract
(NAS8-40214). D.K. was supported by a travel grant to the DSAZ from
the DFG.  J.W.K. was supported by NASA contract NAS5-27000.


\begin{references}
\reference{}
Allard N.F., Kielkopf J.F. 1982, Rev. Modern Phys., 54, 1103
\reference{}
Allard N.F., Kielkopf J.F. 1991, \aap, 242, 133
\reference{}
Allard N.F., Koester D. 1992, \aap, 258, 464
\reference{}
Allard N.F., Koester D., Feautrier N., Spielfiedel A. 1994, A\&AS,
            108, 417
\reference{}
Anderson P.W., Talman J.D. 1956, Bell Telephone Systems Technical
Publications No. 3117, 29 (Murray Hill, NJ)
\reference{}
Bergeron P., Saffer R.A., Liebert J. 1992, \apj, 394, 228
\reference{}
Bergeron P., Wesemael F., Lamontagne R., Fontaine G., Saffer R.,
             Allard N.F. 1995, \apj, 449, 258
\reference{}
Cheselka M., Holberg J.B., Watkins R., Collins J., \& Tweedy R. 1993,
	\aj, 106, 2365
\reference{}
Davidsen A.F. et al. 1992, \apj, 392, 264
\reference{}
Finley D.S., Basri G., \& Bowyer S. 1990, \apj, 359, 483
\reference{}
Finley D.S., Koester D., Basri G. 1996, in preparation
\reference{}
Grewing M., Kraemer G., Appenzeller I. 1991, in Extreme Ultraviolet
Astronomy, eds. R. Malina \& S. Bowyer, Pergamon Press, p. 437
\reference{}
Holweger H., Koester D., Allard N. F. 1994, \aap, 290, L21
\reference{}
Hurwitz M., Bowyer S. 1991, in Extreme Ultraviolet
Astronomy, eds. R. Malina \& S. Bowyer, Pergamon Press, p. 442
\reference{}
Kidder K.M. 1991, PhD Thesis, University of Arizona
\reference{}
Kielkopf J.F., Allard N.F. 1995, \apjl, 450, L75
\reference{}
Koester D., Allard N.F. 1993, in White Dwarfs: Advances in
Observation and Theory, ed. M. Barstow (Kluwer: Dordrecht) p. 237
\reference{}
Koester D., Allard N.F., Vauclair, G. 1994, \aap, 291, L9
\reference{}
Koester D., Weidemann V., Zeidler-K.T. E.-M., Vauclair G. 1985a, \aap
     142, L5
\reference{}
Koester D., Schulz H., Weidemann V. 1979, \aap, 76, 262
\reference{}
Kruk J.W., Durrance S.T., Kriss G.A., Davidsen A.F., Blair W.P.,
Espey B.R.,  Finley D.S. 1995, \apjl, 454, L1
\reference{}
Madsen M.M., Peek J.M. 1971, Atomic Data, 2, 171
\reference{}
Nelan E.P., Wegner G. 1985, \apj, 289, L31
\reference{}
Ramaker D.E., Peek J.M. 1972, J.Phys. B, 5, 2175
\reference{}
Underhill A.B., Waddell J.H. 1959, NBS Circular 603
\end{references}
\end{document}